\documentclass[twocolumn,showpacs,preprintnumbers]{revtex4}
\usepackage{graphicx}
\usepackage{dcolumn}
\usepackage{bm}

\input{tcilatex}

\begin{document}

\title{Valence bond spin liquid state in two-dimensional frustrated spin-1/2
Heisenberg antiferromagnets}
\author{Guang-Ming Zhang$^1$, Hui Hu$^2$, and Lu Yu$^3$}
\affiliation{$^1$Center for Advanced Study, Tsinghua University, Beijing 100084, China;\\
$^2$Abdus Salam International Center for Theoretical Physics, P.O.Box 586,
Trieste 34100, Italy;\\
$^3$Interdisciplinary Center of Theoretical Studies and Institute of
Theoretical Physics, CAS, Beijing 100080, China.}
\date{\today}

\begin{abstract}
Fermionic valence bond approach in terms of SU(4) representation is proposed
to describe the $J_{1}-J_{2}$ frustrated Heisenberg antiferromagnetic (AF)
model on a \textit{bipartite} square lattice. A uniform mean field solution
without breaking the translational and rotational symmetries describes a
valence bond spin liquid state, interpolating the two different AF ordered
states in the large $J_{1}$ and large $J_{2}$ limits, respectively. This
novel spin liquid state is gapless with the vanishing density of states at
the Fermi nodal points. Moreover, a sharp resonance peak in the dynamic
structure factor is predicted for momenta $\mathbf{q}=(0,0)$ and $(\pi ,\pi )
$ in the strongly frustrated limit $J_{2}/J_{1}\sim 1/2$, which can be
checked by neutron scattering experiment.
\end{abstract}

\maketitle

A spin liquid ground state without any symmetry breaking is regarded as one
of the most fascinating possibilities allowed by the physics of
two-dimensional quantum antiferromagnets.\cite{anderson87} It is argued that
reduced dimensionality, a small spin value, and the presence of competing
interactions may lead to strong enough quantum fluctuations to destroy
magnetic long-range order (LRO). A realistic prototype is the quantum
two-dimensional spin-1/2 Heisenberg model with the nearest neighbor (NN) and
next-nearest neighbor (NNN) antiferromagnetic (AF) couplings,\cite{chandra88}
which has been recently materialized experimentally in Li$_{2}$VOSiO$_{4}$
and Li$_{2}$VOGeO$_{4}$ compounds.\cite{melzi} The model is defined on a
square lattice by 
\begin{equation}
H=J_{1}\sum_{n.n}\mathbf{S}_{i}\cdot \mathbf{S}_{j}+J_{2}\sum_{n.n.n}\mathbf{%
S}_{i}\cdot \mathbf{S}_{j},
\end{equation}%
where $J_{1}$, $J_{2}>0$, and the AF alignment between spins on the NN sites
is hindered by the AF coupling of spins between NNN sites.

It is known that, when $J_{2}/J_{1}\ll 1/2$, the ground state has the
conventional N\'{e}el LRO with magnetic wave vector $\mathbf{Q}=(\pi ,\pi )$%
, while in the opposite limit $1>J_{2}/J_{1}\gg 1/2$, the minimum energy
corresponds to a collinear state with $\mathbf{Q=}(\pm \pi ,0)$ and $(0,\pm
\pi )$, consisting of two interpenetrating N\'{e}el sublattices with
independent staggered magnetizations.\cite{chandra90} However, when $%
J_{2}/J_{1}\sim 1/2$, \textit{i.e.}, in the strongly frustrated limit, the
degeneracy of the ground state is large, and there is a consensus that it
corresponds to a spin liquid state without LRO.\cite{katov00} What is the
exact nature of this non-magnetic ground state turns out to be one of the
most challenging problems for the physics of frustrated spin systems. There
have been a number of different proposals, including the columnal, or
spin-Peierls state, where the spin rotational symmetry is preserved but the
translational symmetry is broken,\cite{read-sachdev,dagotto,gelfand} the
plaquette state, recovering the x-y symmetry,\cite{zhitomirsky96} a
chiral-spin state, breaking the PT symmetry,\cite{kalmeyer87} or a truely
homogeneous state, not breaking any translational and rotational symmetries.%
\cite{figueirido,wen} For some time the spin-Peierls and plaquette states
seemed to be favored by numerical studies,\cite{singh,capriotti00,jongh00}
but the most recent numerical simulations \cite{sorella01,imada} on finite
size lattices show strong evidence against all states breaking translational
symmetry, including spin-Peierls and plaquette states, staggered flux phase,
etc. Moreover, Capriotti \textit{et. al., }\cite{sorella01} have argued that
the corresponding strongly frustrated ground state may be characterized by a
projective BCS-type wave function though the two different AF LRO states in
large $J_{1}$ and large $J_{2}$ limits failed to be reproduced. In view of
this latest development the earlier proposal of homogeneous spin liquid state%
\cite{figueirido,wen} appears now as a promising candidate. How to construct
such a state, and how can it interpolate the two different AF LRO states in
two opposite limits is an outstanding issue.

In this Letter, by using an SU(4) constrained fermion representation to
describe the spin-1/2 operators on both sublattices simultaneously, we
propose a fermionic valence bond (VB) approach to describe the $J_{1}-J_{2}$
frustrated Heisenberg AF model on a \textit{bipartite} square lattice. A
uniform mean field (MF) solution gives rise to a new VB spin liquid state
with gapless spin excitations. At zero temperature, due to the inter-band
nesting between two quasiparticle dispersions, the dynamic spin structure
factor (DSSF) at momenta $\mathbf{q=}(0,0)$ and $(\pi ,\pi )$ for spins on
the same sublattice and on different sublattices are found to be equal to
each other up to sign, and a sharp resonance peak is formed in the strongly
frustrated limit $J_{2}/J_{1}\sim 0.5$. Although the \textit{true} AF LRO
can not be reproduced in the weakly frustrated limits, a broad peak at $(\pi
,\pi )$ for $J_{2}/J_{1}\ll 0.5$ is dominant in the static spin structure
factor (SSSF) of spins on \textit{different} sublattices, while for $%
J_{2}/J_{1}\gg 0.5$ rather sharp peaks appear at $(0,\pm \pi )$ and $(\pm
\pi ,0)$ for spins on the \textit{same} sublattice. It has thus been
demonstrated a smooth crossover from the N\'{e}el to collinear AF (quasi)
LRO when changing $J_{2}/J_{1}$ from small to large values, and a uniform VB
spin liquid phase interpolates between the two weakly frustrated regimes.

The $J_{1}-J_{2}$ model on a \textit{bipartite} square lattice can be
rewritten as 
\begin{eqnarray}
H &=&\frac{J_{1}}{2}\sum_{<ij>}\left( \mathbf{S}_{i,A}\cdot \mathbf{S}_{j,B}+%
\mathbf{S}_{i,B}\cdot \mathbf{S}_{j,A}\right)  \nonumber \\
&&+J_{2}\sum_{(i.l)}\left( \mathbf{S}_{i,A}\cdot \mathbf{S}_{l,A}+\mathbf{S}%
_{i,B}\cdot \mathbf{S}_{l,B}\right) ,
\end{eqnarray}%
where $<ij>$ denotes summation over NN belonging to \textit{different}
sublattices, while $(i,l)$ means summation over the NN sites of the \textit{%
same} sublattice. Each sublattice has $N$ sites. As we know, the bipartite
lattice structure is essential for the spin density wave (SDW) theory in
describing AF LRO state, where the correspondence between the spin operators
on the two sublattices is assumed as $S_{i,B}^{+}\rightarrow S_{i,A}^{-}$, $%
S_{i,B}^{-}\rightarrow S_{i,A}^{+}$, and $S_{i,B}^{z}\rightarrow
-S_{i,A}^{z} $, reflecting the presence of the N\'{e}el LRO. Actually, the
bipartite lattice structure is also an important setting to study various
quantum magnetic systems even in the \textit{absence} of AF LRO. However,
the correspondence between the spin operators on the two sublattices in the
VB state is no longer the same as that in the SDW theory.

We generalize the conventional SU(2) constrained fermion to an SU(4)
representation \cite{gmzhang}. The generators of the SU(4) fermion
representation are given by $F_{\beta }^{\alpha }(i)=C_{i,\alpha }^{\dagger
}C_{i,\beta }$, satisfying the SU(4) Lie algebra: $[F_{\beta }^{\alpha
}(i),F_{\nu }^{\mu }(j)]=[\delta _{\beta ,\mu }F_{\nu }^{\alpha }(i)-\delta
_{\alpha ,\nu }F_{\beta }^{\mu }(i)]\delta _{i,j}$. The spin operators on
the sublattice $A$ can be expressed as 
\begin{eqnarray*}
S_{i,A}^{+} &=&C_{i,1}^{\dagger }C_{i,2}+C_{i,3}^{\dagger }C_{i,4}, \\
S_{i,A}^{-} &=&C_{i,2}^{\dagger }C_{i,1}+C_{i,4}^{\dagger }C_{i,3}, \\
S_{i,A}^{z} &=&\frac{1}{2}\left[ (C_{i,1}^{\dagger }C_{i,1}-C_{i,2}^{\dagger
}C_{i,2})+(C_{i,3}^{\dagger }C_{i,3}-C_{i,4}^{\dagger }C_{i,4})\right] ,
\end{eqnarray*}%
while $S_{i,B}^{\pm }$ and $S_{i,B}^{z}$ are given by interchanging $%
C_{i,2}\longleftrightarrow C_{i,3}$, reflecting the symmetry of the model.
With this representation, $S_{i,A}^{\alpha }$ and $S_{i,B}^{\alpha }$ ($%
\alpha =x,y,z$) are proved to satisfy their respective commutation relations
of the SU(2) Lie algebra and $[S_{i,A}^{\alpha },S_{i,B}^{\beta }]=0$. By
imposing a local constraint $\sum_{\mu }C_{i,\mu }^{\dagger }C_{i,\mu }=1$
on each lattice site, we can further prove that $\mathbf{S}_{i,A}^{2}=%
\mathbf{S}_{i,B}^{2}=3/4$.

In this new representation, the $J_1-J_2$ model can be rewritten as 
\begin{eqnarray}
H &=&-\frac{J_1}4\sum_{<ij>}\left( :A_{i,j}^{\dagger
}A_{i,j}:+B_{i,j}^{\dagger }B_{i,j}\right)  \nonumber \\
&&-\frac{J_2}2\sum_{(i,l)}\left( :P_{i,l}^{\dagger
}P_{i,l}:+Q_{i,l}^{\dagger }Q_{i,l}\right) ,
\end{eqnarray}
where the normal ordering has been taken for the first and third terms, and
four composite VB order operators have been introduced 
\begin{eqnarray*}
A_{i,j} &=&[(C_{j,1}^{\dagger }C_{i,1}+C_{j,4}^{\dagger
}C_{i,4})+(C_{j,3}^{\dagger }C_{i,2}+C_{j,2}^{\dagger }C_{i,3})], \\
P_{i,l} &=&[(C_{l,1}^{\dagger }C_{i,1}+C_{l,4}^{\dagger
}C_{i,4})+(C_{l,2}^{\dagger }C_{i,2}+C_{l,3}^{\dagger }C_{i,3})], \\
B_{i,j} &=&\left[ \left( C_{j,4}C_{i,1}+C_{j,1}C_{i,4}\right) -\left(
C_{j,2}C_{i,2}+C_{j,3}C_{i,3}\right) \right] , \\
Q_{i,l} &=&\left[ \left( C_{l,4}C_{i,1}+C_{l,1}C_{i,4}\right) -\left(
C_{l,3}C_{i,2}+C_{l,2}C_{i,3}\right) \right] .
\end{eqnarray*}

When uniform VB order parameters are assumed that $\langle A_{i,j}\rangle
=\Delta _{1c}$, $\langle B_{i,j}\rangle =-\Delta _{1s}$, $\langle
P_{i,l}\rangle =\Delta _{2c}$, $\langle Q_{i,l}\rangle =-\Delta _{2s}$, and
the local constraint is replaced by a uniform Lagrangian multiplier, the MF
model Hamiltonian is obtained as 
\begin{eqnarray}
H_{mf} &=&\frac{1}{2}\sum_{\mathbf{k}}\Psi _{\mathbf{k}}^{\dagger }\left\{ %
\left[ \lambda -\Delta _{1c}(\mathbf{k})-\Delta _{2c}(\mathbf{k})\right] 
\mathbf{\Omega }_{1}\right.  \nonumber \\
&&\left. -\left[ \Delta _{1s}(\mathbf{k})+\Delta _{2s}(\mathbf{k})\right] 
\mathbf{\Omega }_{2}\right\} \Psi _{\mathbf{k}}  \nonumber \\
&&+\frac{1}{2}\sum_{\mathbf{k}}\Phi _{\mathbf{k}}^{\dagger }\left\{ \left[
\lambda -\Delta _{2c}(\mathbf{k})\right] \mathbf{\Omega }_{1}-\Delta _{1c}(%
\mathbf{k})\mathbf{\Omega }_{3}\right.  \nonumber \\
&&\left. +\Delta _{1s}(\mathbf{k})\mathbf{\Omega }_{4}+\Delta _{2s}(\mathbf{k%
})\mathbf{\Omega }_{2}\right\} \Phi _{\mathbf{k}}  \nonumber \\
&&+2N\left[ J_{1}\left( \Delta _{1c}^{2}+\Delta _{1s}^{2}\right)
+J_{2}\left( \Delta _{2c}^{2}+\Delta _{2s}^{2}\right) +\lambda \right], \nonumber
\end{eqnarray}%
where the Nambu spinors are introduced as $\Psi _{\mathbf{k}}^{\dagger }=(C_{%
\mathbf{k},1}^{\dagger },C_{\mathbf{k},4}^{\dagger },C_{-\mathbf{k},1},C_{-%
\mathbf{k},4})$, $\Phi _{\mathbf{k}}^{\dagger }=(C_{\mathbf{k},2}^{\dagger
},C_{\mathbf{k},3}^{\dagger },C_{-\mathbf{k},2},C_{-\mathbf{k},3})$, and $%
4\times 4$ matrices are defined by $\mathbf{\Omega }_{1}=\sigma _{z}\otimes
\sigma _{0}$, $\mathbf{\Omega }_{2}=\sigma _{y}\otimes \sigma _{x}$, $%
\mathbf{\Omega }_{3}=\sigma _{z}\otimes \sigma _{x}$, $\mathbf{\Omega }%
_{4}=\sigma _{y}\otimes \sigma _{0}$. Moreover, $\Delta _{1c}(\mathbf{k}%
)=J_{1}\Delta _{1c}(\cos k_{x}+\cos k_{y})$, $\Delta _{2c}(\mathbf{k}%
)=4J_{2}\Delta _{2c}\cos k_{x}\cos k_{y}$, $\Delta _{1s}(\mathbf{k}%
)=J_{1}\Delta _{1s}\left( \sin k_{x}+\sin k_{y}\right) $, and $\Delta _{2s}(%
\mathbf{k})=4J_{2}\Delta _{2s}\sin k_{x}\cos k_{y}$. The fermionic Matsubara
Green function matrices are then derived as 
\begin{eqnarray*}
\mathbf{G}_{1,4}(\mathbf{k},i\omega _{n}) &=&\left[ i\omega _{n}-\left(
\lambda -\Delta _{1c}(\mathbf{k})-\Delta _{2c}(\mathbf{k})\right) \mathbf{%
\Omega }_{1}\right. \\
&&\left. +\left( \Delta _{1s}(\mathbf{k})+\Delta _{2s}(\mathbf{k})\right) 
\mathbf{\Omega }_{2}\right] ^{-1}, \\
\mathbf{G}_{2,3}(\mathbf{k},i\omega _{n}) &=&\left[ i\omega _{n}-\left(
\lambda -\Delta _{2c}(\mathbf{k})\right) \mathbf{\Omega }_{1}+\Delta _{1c}(%
\mathbf{k})\Omega _{3}\right. \\
&&\left. -\Delta _{1s}(\mathbf{k})\Omega _{4}-\Delta _{2s}(\mathbf{k})%
\mathbf{\Omega }_{2}\right] ^{-1},
\end{eqnarray*}%
whose poles give rise to the quasiparticle dispersions 
\[
\epsilon _{\pm }(\mathbf{k})=\sqrt{\left( \lambda \mp \Delta _{1c}(\mathbf{k}%
)-\Delta _{2c}(\mathbf{k})\right) ^{2}+\left( \Delta _{1s}(\mathbf{k})\pm
\Delta _{2s}(\mathbf{k})\right) ^{2}} 
\]%
where $\epsilon _{+}(\mathbf{k})$ corresponds to the triplet excitations
with three-fold degeneracy, and $\epsilon _{-}(\mathbf{k})$ corresponds to
the singlet excitations. However, an \textit{inter-band} nesting property,
namely $\epsilon _{+}(\mathbf{k})=\epsilon _{-}(\mathbf{k+Q})$ with a
nesting wave vector $\mathbf{Q}=(\pi ,\pi )$ is a very important feature for
the two quasiparticle bands. The usual intra-band nesting for the
half-filled Hubbard model leads to the AF LRO at zero temperature.
Similarly, the inter-band nesting will make the AF quasi-long range
correlations dominant, excluding any possible incommensurate density wave
states.

Hence the ground state energy per site is obtained and simplified as 
\[
\varepsilon =-\frac{1}{N}\sum_{\mathbf{k}}\epsilon _{+}(\mathbf{k}%
)+J_{1}\left( \Delta _{1c}^{2}+\Delta _{1s}^{2}\right) +J_{2}\left( \Delta
_{2c}^{2}+\Delta _{2s}^{2}\right) +\lambda . 
\]%
The saddle point equations are derived by minimizing the ground state energy
with respect to $\Delta _{1c}$, $\Delta _{2c}$, $\Delta _{1s}$, $\Delta
_{2s} $, and $\lambda $. 
By solving those equations, we can determine the saddle point parameters as
functions of $J_{2}/J_{1}$. We notice that the two pairing order parameters
are competing with each other: when $J_{2}/J_{1}<0.545$, $\Delta
_{1s}>\Delta _{2s}$, while for $J_{2}/J_{1}\geq 0.545$, $\Delta _{1s}\leq
\Delta _{2s}$.

Due to the inter-band nesting property, only the quasiparticle spectrum $%
\epsilon _{+}(\mathbf{k})$ is considered for different coupling parameters $%
J_{2}/J_{1}$. Although the uniform VB order parameters are assumed above, $%
\epsilon _{+}(\mathbf{k})$ displays two nodal Fermi points at $(\mp
k_{0,x},\pm k_{0,y})$, and the quasiparticle density of states \textit{%
algebraically} vanishes at the Fermi points. Moreover, when $J_{2}/J_{1}\ll
0.5$, the two nodal Fermi points are very close to the diagonal line $%
k_{x}+k_{y}=0$, while for $J_{2}/J_{1}\gg 0.5$ the two nodal Fermi points
are rotated versus the vertical line $k_{x}=0$. The position of one nodal
Fermi point $(-k_{0,x},k_{0,y})$ is plotted as a function of $J_{2}/J_{1}$
in Fig.1.
\begin{figure}[tbp]
\begin{center}
\includegraphics[width=2.7in]{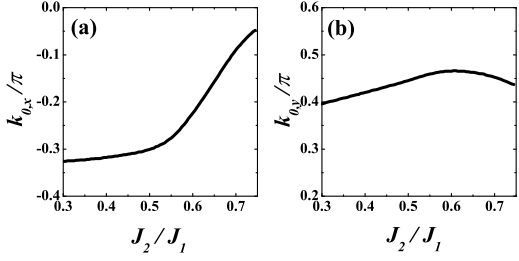}
\end{center}
\caption{The position of the nodal Fermi point $(-k_{0,x},k_{0,y})$ in the
quasiparticle spectrum $\protect\epsilon _{-}(\mathbf{k})$ as a function of $%
J_{2}/J_{1}$. Due to the inversion symmetry, another nodal point is at $%
(k_{0,x},-k_{y})$.}
\end{figure}

In order to reveal the nature of the new VB spin liquid state, the DSSFs are
calculated. In terms of Nambu spinors $\Psi _{i}$ and $\Phi _{i}$, the spin
operators on the two sublattices are written as $S_{i,A}^{z}=(\Psi
_{i}^{\dagger }\mathbf{\Omega }_{5}\Psi _{i,A}-\Phi _{i}^{\dagger }\mathbf{%
\Omega }_{5}\Phi _{i})/4$ and $S_{i,B}^{z}=(\Psi _{i}^{\dagger }\mathbf{%
\Omega }_{5}\Psi _{i,A}+\Phi _{i}^{\dagger }\mathbf{\Omega }_{5}\Phi _{i})/4$%
, where $\mathbf{\Omega }_{5}=\sigma _{z}\otimes \sigma _{z}$, and in the
Fourier space the DSSF for spins on the same and different sublattices are
given by,%
\begin{equation}
\left( 
\begin{array}{c}
{\chi _{A,A}^{z}(\mathbf{q,}i\omega _{m})} \\ 
{\chi _{A,B}^{z}(\mathbf{q,}i\omega _{m})}%
\end{array}%
\right) =\chi _{1,4}(\mathbf{q,}i\omega _{m})\pm \chi _{2,3}(\mathbf{q,}%
i\omega _{m}),
\end{equation}%
where 
\begin{eqnarray}
&&\chi _{1,4}(\mathbf{q,}i\omega _{m})=-\frac{1}{16\beta }\sum_{\omega
_{n}}\int \frac{d^{2}\mathbf{k}}{(2\pi )^{2}}  \nonumber \\
&&\hspace{1cm}\mathrm{Tr}\left[ \mathbf{\Omega }_{5}\mathbf{G}_{1,4}(\mathbf{%
k},i\omega _{n})\mathbf{\Omega }_{5}\mathbf{G}_{1,4}(\mathbf{q+k},i\omega
_{m}+i\omega _{n})\right], \nonumber
\end{eqnarray}%
and $\chi _{2,3}(\mathbf{q,}i\omega _{m})$ can be obtained by replacing $%
\mathbf{G}_{1,4}$ with $\mathbf{G}_{2,3}$ in $\chi _{1,4}(\mathbf{q,}i\omega
_{m})$. Due to the vanishing density of states of quasiparticles around the
nodal Fermi points, the DSSFs are not divergent at zero temperature,
yielding only algebraically decaying spin-spin correlations (quasi-LRO).

By considering the inter-band nesting, the following important relations can
be further proved 
\begin{eqnarray}
&&\chi _{A,A}^{z}(\mathbf{Q,}i\omega _{m})=\chi _{A,B}^{z}(\mathbf{Q,}%
i\omega _{m})  \nonumber \\
&=&\chi _{A,A}^{z}(\mathbf{0,}i\omega _{m})=-\chi _{A,B}^{z}(\mathbf{0,}%
i\omega _{m})\equiv \chi (\mathbf{Q,}i\omega _{m}),
\end{eqnarray}%
giving rise to the characteristic features of frustrated AF correlations of
spins on the bipartite square lattice! After analytic continuation, its
imaginary part of the correlation spectrum at $T=0$ is approximated to be 
\begin{eqnarray*}
&&\text{Im}\chi (\mathbf{Q,}\omega >0)\approx \frac{\pi }{16}\int \frac{d^{2}%
\mathbf{k}}{(2\pi )^{2}}\delta \left[ \omega -\epsilon _{+}(\mathbf{k}%
)-\epsilon _{-}(\mathbf{k})\right] \\
&&\hspace{1cm}\left\{ 1-\frac{\left[ \lambda -\Delta _{2c}(\mathbf{k})\right]
^{2}-\Delta _{1c}^{2}(\mathbf{k})-\Delta _{1s}^{2}(\mathbf{k})+\Delta
_{2s}^{2}(\mathbf{k})}{\epsilon _{+}(\mathbf{k})\epsilon _{-}(\mathbf{k)}}%
\right\} ,
\end{eqnarray*}%
which is delineated in Fig.2. Increasing the coupling parameter $J_{2}/J_{1}$%
, a sharp resonance is gradually formed in the strongly frustrated limit $%
J_{2}/J_{1}\sim 0.5$ and disappears quickly away from it. The resonance peak
simultaneously appears at momenta $(\pi ,\pi )$ and $(0,0)$, and may be
regarded as a signature of the present fermionic VB spin liquid state in the
strongly frustrated limit of the model, which is ready to be verified by
future experiments.

\begin{figure}[tbp]
\begin{center}
\includegraphics[width=2.7in]{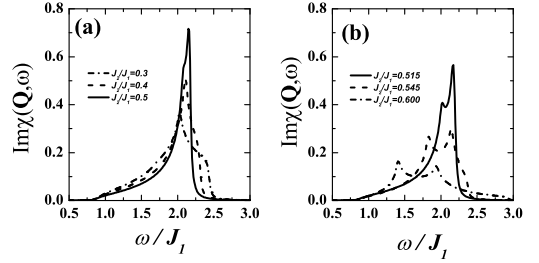}
\end{center}
\caption{The imaginary part of the dynamic spin structure factor ${Im}%
\protect\chi (\mathbf{Q,}\protect\omega >0)$ for different values of $%
J_2/J_1 $.}
\end{figure}

Moreover, to compare with the AF LRO states in the large $J_{1}$ and large $%
J_{2}$ regimes, respectively, the SSSFs at $T=0$ are evaluated. 
A smooth crossover is obtained for the SSSFs $\chi _{AA}^{z}(\mathbf{q})$
and $\chi _{AB}^{z}(\mathbf{q})$ when increasing the coupling parameter
through the strongly frustrated limit $J_{2}/J_{1}=0.5$. In Fig.3, for $%
J_{2}/J_{1}=0.3$ a rather broad peak is clearly identified at $\mathbf{q}%
=(\pi ,\pi )$ in $\chi _{A,B}^{z}(\mathbf{q})$, while $\chi _{A,A}^{z}(%
\mathbf{q})$ has no distinctive features. This peak manifests the dominant
quasi-long range AF correlations towards the N\'{e}el LRO in the limit of
small $J_{2}/J_{1}$. In contrary, in Fig.4 for $J_{2}/J_{1}=0.7$, the SSSF $%
\chi _{A,A}^{z}(\mathbf{q})$ displays rather sharp peaks at momenta $(\pm
\pi ,0)$ and $(0,\pm \pi )$, while the SSSF $\chi _{A,B}^{z}(\mathbf{q})$ is
very small and spreads over a certain range. These peculiar features
indicate the quasi-long range AF correlations towards the collinear LRO in
the limit of large $J_{2}/J_{1}$. Therefore, it has been shown that a
fermionic VB spin liquid state in terms of the SU(4) representation exists
and approximately interpolates between the two AF LRO states in the large $%
J_{1}$ and large $J_{2}$ limits, respectively.

\begin{figure}[tbp]
\begin{center}
\includegraphics[width=2.7in]{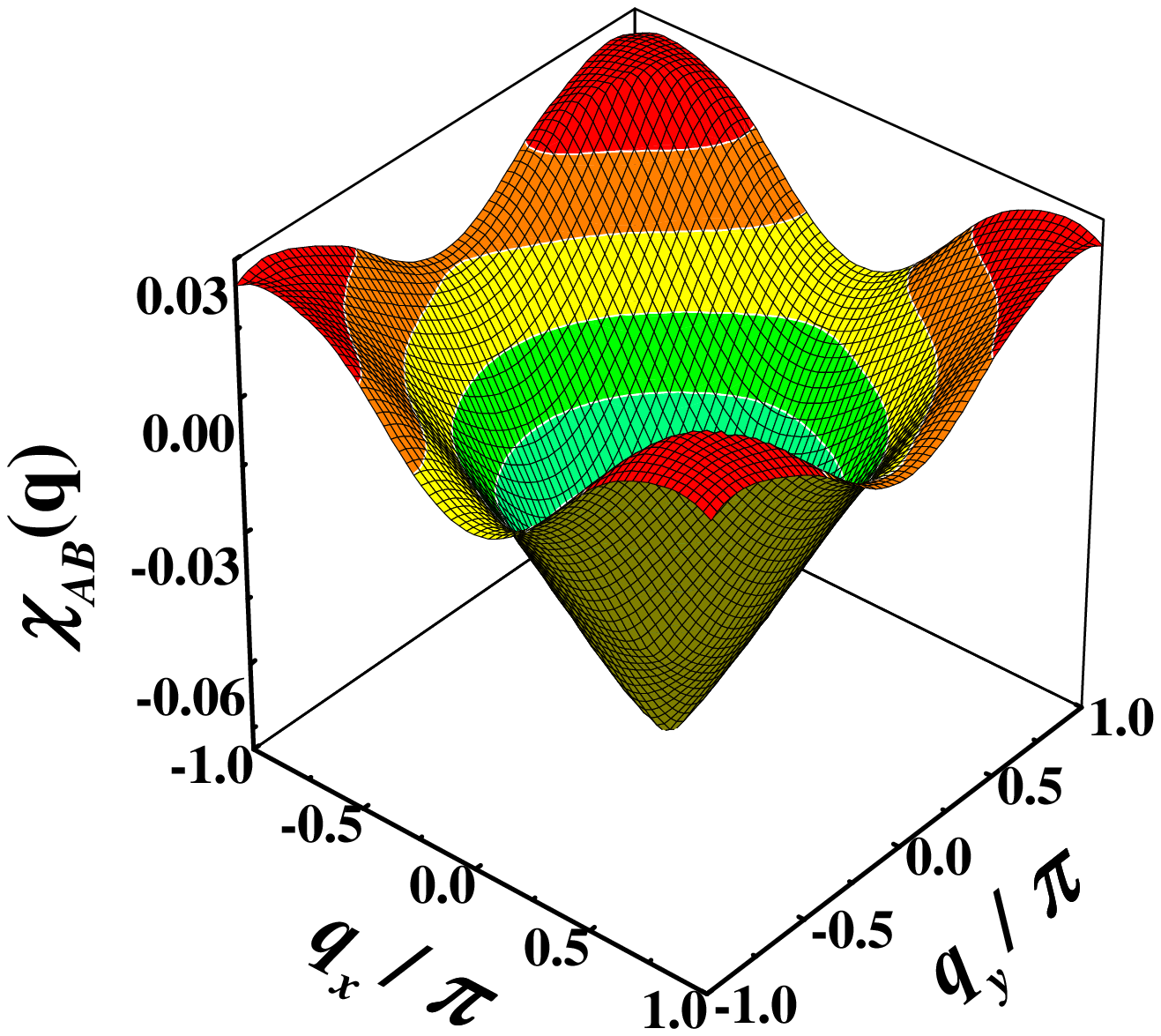}
\end{center}
\caption{The static spin structure factor at zero temperature $\protect\chi %
_{A,B}(\mathbf{q})$ for $J_2/J_1=0.3$. A broad peak at $\left( \protect\pi ,%
\protect\pi \right) $ displays quasi-long range AF correlations in the large 
$J_1$ limit.}
\end{figure}

\begin{figure}[tbp]
\begin{center}
\includegraphics[width=2.7in]{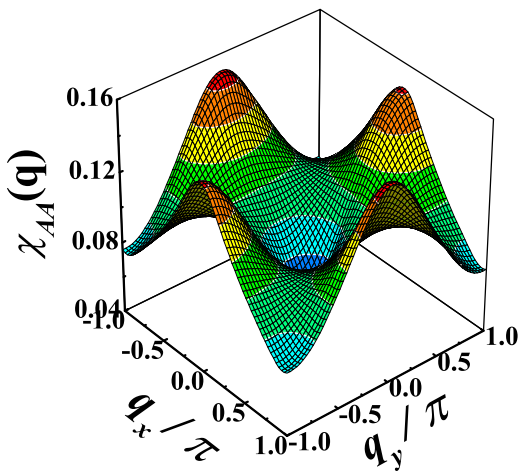}
\end{center}
\caption{The static spin structure factor at zero temperature $\protect\chi %
_{A,A}(\mathbf{q})$ for $J_2/J_1=0.7$. Sharp peaks appearing at $\left( \pm 
\protect\pi ,0\right) $ and $(0,\pm \protect\pi )$ indicate the strong
quasi-long range correlations of the collinear AF state.}
\end{figure}

In conclusion, an SU(4) constrained fermion representation has been used to
describe the $J_{1}-J_{2}$ frustrated Heisenberg AF model on a bipartite
square lattice, and a fermionic uniform MF solution gives rise to a novel VB
spin liquid state with gapless spin excitations, not breaking any
translational and rotational symmetries. This MF solution has reproduced
correct asymptotic behavior in the weakly frustrated limits with the
provision that the true LRO is approximated by quasi-LRO in our scheme. To
the best of our knowledge, no other theoretical approach has succeeded in
doing so in this complicated problem. The success of our MF theory is mainly
due to the appropriate choice of representation which grasps the essential
features of the strongly frustrated state. Of course, the eventual ``proof''
of our theory should come from the experiment. In particular, our explicit
prediction of the sharp resonance peak at momenta $\left( \pi ,\pi \right) $
and $(0,0)$ in the imaginary part of the DSSF in the strongly frustrated
limit $J_{2}/J_{1}\sim 0.5$ can serve as a crucial experimental test. There
are still many open problems, for example, what is the nature of transitions
between this VB state and AF LRO states, their possible coexistence, etc.
These issues require further studies.

We are grateful to Professor Z. B. Su for useful discussions. This work is
supported by NSF-China (Grant Nos. 10074036 and 10125418) and the Special
Fund for Major State Basic Research Projects of China (Grant No.
G2000067107).

\end{document}